\documentclass[letter,conference]{IEEEtran}

\usepackage{csvsimple}
\usepackage{threeparttable}
\usepackage{subfigure}

\usepackage{tikz}
\newcommand*\circled[1]{\tikz[baseline=(char.base)]{
    \node[shape=circle,draw,inner sep=2pt] (char) {#1};}}

\usepackage{pgfplots}
\usepgfplotslibrary{colorbrewer}

\usepackage{hyperref}
\usepackage[square,numbers]{natbib}
\usepackage{graphicx}
\usepackage{textcomp}
\usepackage{xcolor}
\usepackage{booktabs}

\usepackage{amsmath}
\usepackage{amsfonts}
\usepackage{amssymb}
\usepackage{multirow}

\usepackage[ruled,vlined,linesnumbered]{algorithm2e}
\DontPrintSemicolon
\SetArgSty{}
\SetKw{Continue}{continue}
\SetKw{Break}{break}
\SetKw{Print}{print}

\usepackage{listings}

\definecolor{mediumgray}{rgb}{0.3, 0.4, 0.4}
\definecolor{mediumblue}{rgb}{0.0, 0.0, 0.8}
\definecolor{forestgreen}{rgb}{0.13, 0.55, 0.13}
\definecolor{darkviolet}{rgb}{0.58, 0.0, 0.83}
\definecolor{royalblue}{rgb}{0.25, 0.41, 0.88}
\definecolor{crimson}{rgb}{0.86, 0.8, 0.24}

\lstdefinestyle{JSES6Base}{
	backgroundcolor=\color{white},
	basicstyle=\ttfamily,
	breakatwhitespace=false,
	breaklines=false,
	captionpos=b,
	columns=fullflexible,
	commentstyle=\color{mediumgray}\upshape,
	emph={},
	emphstyle=\color{crimson},
	extendedchars=true,  
	firstnumber=1,
	fontadjust=true,
	frame=none,
	identifierstyle=\color{black},
	keepspaces=true,
	keywordstyle=\color{mediumblue},
	keywordstyle={[2]\color{darkviolet}},
	keywordstyle={[3]\color{royalblue}},
	numbers=left,
	numbersep=5pt,
	numberstyle=\tiny\color{black},
	rulecolor=\color{black},
	showlines=true,
	showspaces=false,
	showstringspaces=false,
	showtabs=false,
	stringstyle=\color{forestgreen},
	tabsize=2,
	title=\lstname,
	upquote=true,  
	xleftmargin=1em,
        basicstyle=\footnotesize\ttfamily
}

\lstdefinelanguage{JavaScript}{
	morekeywords=[1]{break, continue, else, for, function, if, 
		new, return, this, typeof, var, void, while, with, await, async, case, 
		catch, class, const, default, do, enum, export, extends, finally, from, 
		implements, import, instanceof, let, static, super, switch, throw, try},
	morekeywords=[2]{false, null, true, boolean, number, undefined,
		Array, Boolean, Date, Math, Number, String, Object},
	morekeywords=[3]{eval, parseInt, parseFloat, escape, unescape},
	sensitive,
	morecomment=[s]{/*}{*/},
	morecomment=[l]//,
	morecomment=[s]{/**}{*/}, 
	morestring=[b]',
	morestring=[b]",
	morestring=[b]` 
}[keywords, comments, strings]

\lstdefinelanguage{json}
{
	morestring=[b]",
	morestring=[d]'
}

\lstdefinestyle{JavaScript}{
	language=JavaScript,
	style=JSES6Base
}

\newcommand{\codett}[1]{\lstinline[style=javascript]{#1}}

\newif\ifnoblind

\noblindtrue

\makeatletter
\newcommand{\linebreakand}{%
\end{@IEEEauthorhalign}
\hfill\mbox{}\par
\mbox{}\hfill\begin{@IEEEauthorhalign}
}
\makeatother

\begin{document}

\title{BackREST: A Model-Based Feedback-Driven Greybox Fuzzer for Web Applications}

\author{\IEEEauthorblockN{1\textsuperscript{st} Fran\c{c}ois Gauthier}
\IEEEauthorblockA{\textit{Oracle Labs} \\
Brisbane, Australia \\
francois.gauthier@oracle.com}
\and
\IEEEauthorblockN{2\textsuperscript{nd} Behnaz Hassanshahi}
\IEEEauthorblockA{\textit{Oracle Labs} \\
Brisbane, Australia \\
behnaz.hassanshahi@oracle.com}
\and
\IEEEauthorblockN{3\textsuperscript{rd} Benjamin Selwyn-Smith}
\IEEEauthorblockA{\textit{Oracle Labs} \\
Brisbane, Australia \\s
benselwynsmith@googlemail.com}
\linebreakand
\IEEEauthorblockN{4\textsuperscript{th} Trong Nhan Mai}
\IEEEauthorblockA{\textit{Oracle Labs} \\
Brisbane, Australia \\
trongnhan.mai@uqconnect.edu.au}
\and
\IEEEauthorblockN{5\textsuperscript{th}Max Schl{\" u}ter}
\IEEEauthorblockA{\textit{Oracle Labs} \\
Brisbane, Australia \\
mschlueter@uni-potsdam.de}
\and 
\IEEEauthorblockN{6\textsuperscript{th} Micah Williams}
\IEEEauthorblockA{\textit{Oracle} \\
Morrisville, NC, USA \\
micah.williams@oracle.com}
}

\maketitle

\begin{abstract}
Following the advent of the American Fuzzy Lop (AFL), fuzzing had a surge in popularity, 
and modern day fuzzers range from simple blackbox random input generators to complex 
whitebox concolic frameworks that are capable of deep program introspection. Web 
application fuzzers, however, did not benefit from the tremendous advancements in
fuzzing for binary programs and remain largely blackbox in nature. This paper introduces 
\textsc{BackREST}, a fully automated, model-based, coverage- and taint-driven fuzzer that 
uses its feedback loops to find \emph{more} critical vulnerabilities, \emph{faster} 
(speedups between 7.4$\times$ and 25.9$\times$). To model the server-side of web applications, 
\textsc{BackREST} automatically infers REST specifications through
directed state-aware crawling. Comparing \textsc{BackREST} against three other web fuzzers 
on five large ($>$500 KLOC) Node.js applications shows how it consistently achieves comparable
coverage while reporting more vulnerabilities than state-of-the-art. Finally, using \textsc{BackREST}, 
we uncovered nine 0-days, out of which six were not reported by any other fuzzer. All 
the 0-days have been disclosed and most are now public, including two in the highly 
popular Sequelize and Mongodb libraries.
\end{abstract}

\begin{IEEEkeywords}
	fuzzing, greybox, coverage, taint analysis, REST
\end{IEEEkeywords}

\section{Introduction}

Fuzzing encompasses techniques and
tools to automatically identify vulnerabilities in programs by sending
malformed or malicious inputs and monitoring abnormal behaviours. 
Nowadays, fuzzers come in three major shades: blackbox, greybox, and whitebox, according 
to how much visibility they have into the program internals \cite{godefroid2020fuzzing}. 
Greybox fuzzing, which was made popular by the AFL fuzzer, combines blackbox with 
lightweight whitebox techniques, and have proven to be very effective at fuzzing 
programs that operate on binary input \cite{afl,bohme2017coverage,bohme2017directed}. 



Most web application fuzzers that are used in practice are still blackbox \cite{zap,w3af,burp,jaek},
and, despite decades of development, still struggle to automatically detect well studied 
vulnerabilities such as SQLi, and XSS \cite{rahaman2019security}. As a result, security
testing teams have to invest significant manual efforts into building models of the 
application and driving the fuzzer to trigger vulnerabilities. To overcome the limitations 
of current blackbox web application fuzzers, and find more vulnerabilities automatically,
new strategies must be investigated. To pave the way for the next generation of practical web 
application fuzzers, this paper first shows how REST models are a suitable abstraction
for web applications and how adding lightweight coverage and taint feedback loops to a 
blackbox fuzzer can significantly improve performance and detection capabilities. Then, it highlights how 
the resulting \textsc{BackREST} greybox fuzzer consistently detects more vulnerabilities
than state-of-the-art. Finally, our evaluation reveals how \textsc{BackREST} found nine
0-days, out of which six were \emph{missed} by all the web application fuzzers we compared 
it against.



\paragraph{REST model inference} Model-based fuzzers, which use a model to impose 
\emph{constraints} on input, dominate the web application fuzzing scene. Existing model-based 
web application fuzzers typically use dynamically captured traffic to derive a base model, 
which can be further enhanced manually \cite{peach,burp,zap}. As with any dynamic analysis,
relying on captured traffic makes the quality of the model dependent on the quality of the 
traffic generator, be it a human being, a test suite, or a crawler. To navigate JavaScript-heavy
applications and trigger a maximum number of endpoints, \textsc{BackREST} directs a state-aware
crawler towards JavaScript functions that trigger server-side interactions.

\paragraph{Feedback-driven} What makes greybox fuzzers so efficient is the feedback loop between the lightweight 
whitebox analysis components, and the blackbox input generator. \textsc{BackREST} is the first web application fuzzer that can 
focus the fuzzing session on those areas of the application that haven't been exercised yet (i.e. by using coverage feedback), and that have a higher chance of
containing security vulnerabilities (i.e. by using taint feedback). Taint feedback 
in \textsc{BackREST} further reports the \emph{type} (e.g. SQLi, XSS, or command injection) of 
potential vulnerabilities, enabling
\textsc{BackREST} to aggressively fuzz a given endpoint with vulnerability-specific payloads, yielding tremendous performance improvements in practice.


\paragraph{Validated in practice} The two main metrics that drive practical adoption of a fuzzer are the number of vulnerabilities it can detect and the time required to discover 
them. Our experiments show how \textsc{BackREST} uncovered nine 0-days in five 
large ($>$ 500KLOC) Node.js applications, and how adding lightweight whitebox 
analyses significantly speeds up (7.4-25.9$\times$ faster) the fuzzing session. In 
other words, the time saved by focusing the fuzzing session on unexplored and 
potentially vulnerable areas of the application largely dwarfs the overhead of the 
instrumentation-based analyses. All 0-days have been disclosed, and most have been
announced as NPM advisories, meaning that developers will be alerted about them
when they update their systems. Four of them have been tagged with high or critical 
severity, and two have been reported against the highly popular \texttt{sequelize} 
(648\,745 weekly downloads) and \texttt{mongodb} (1\,671\,653 weekly downloads) libraries.

\paragraph{Contributions} This paper makes the following contributions:
\begin{itemize}
	\item We show how REST APIs can effectively model the server-side of modern web 
	applications, and quantify how coverage and taint feedback enhance coverage, 
	performance, and vulnerability detection.
	\item We empirically evaluate \textsc{BackREST} on five large ($>$500 KLOC) Node.js 
	(JavaScript) web applications; a platform and language that are notoriously 
	difficult to analyse, and under-represented in current security literature.
	\item We compare \textsc{BackREST} against three state-of-the-art fuzzers 
	(Arachni, Zap, and w3af) and open-source our test harness
	 \ifnoblind\footnote{\url{https://github.com/uqcyber/NodeJSFuzzing}}\else\footnote{Anonymized for submission}\fi.
	\item We show how greybox fuzzing for web applications allows to detect 
	\emph{severe} 0-days that are missed by all the blackbox fuzzers we evaluated.
\end{itemize}

The rest of this paper is structured as follows. Section \ref{sec:model-inference} presents
our novel REST model inference technique. Section \ref{sec:feedback-fuzzing} and \autoref{sec:implementation} detail the \textsc{BackREST} feedback-driven fuzzing algorithm
and implementation, respectively. Section \ref{sec:evaluation} evaluates \textsc{BackREST} in terms
of coverage, performance, and detected vulnerabilities and compares it against state-of-the-art
web application fuzzers. Section \ref{sec:case-studies} 
presents and explains reported 0-days. Sections \ref{sec:relwork} and \ref{sec:conclusion}
present related work and conclude the paper.

\section{REST Model Inference}
\label{sec:model-inference}

Web applications expose entry points in the form of URLs that clients can interact with
via the HTTP protocol. However, the HTTP protocol specifies only how a client and a server
can send and receive information over a network connection, not how to structure
the interactions. Nowadays, representational state transfer (REST) is the \emph{de facto} 
protocol that most modern client-side applications use to communicate with their backend 
server. While the REST protocol was primarily 
aimed at governing interactions with web services, we make the fundamental observation 
that client-server interactions in modern web applications can also be modelled with REST.
Indeed, at its core, REST uses standard HTTP verbs, URLs, and request parameters to define 
and encapsulate client-server interactions, which, from our experience, is also what many
modern web application frameworks do.

Despite the plethora of REST-related tools available, the task of creating an initial
REST specification remains, however, largely manual. While tools exist to convert
captured traffic into a REST specification, the burden of thoroughly exercising the
application or augmenting the specification with missing information is still borne by 
developers. \textsc{BackREST} alleviates this manual effort by extending a state-aware crawler
designed for rich client-side JavaScript applications \cite{hassanshahi2020gelato}.

Modern web applications, and single-page ones in particular, implement complex and highly interactive functionalities on the client side. A recent Stack Overflow developer survey~\cite{survey} shows that web applications are increasingly built using complex 
client-side frameworks, such as AngularJS~\cite{angularjs} and React~\cite{React.js}.
In fact, three out of five applications we evaluate in Section \ref{sec:evaluation} heavily 
use such frameworks. Using a state-aware crawler that can automatically navigate
complex client-side frameworks allows \textsc{BackREST} to discover server-side endpoints 
that can only be triggered through complex JavaScript interactions.

\subsection{Motivating Example}

\autoref{example-endpoint} shows an endpoint definition from a Node.js
Express application. At line 1, \lstinline[style=javascript]{app} refers to the
programmatic web application object, and the \lstinline[style=javascript]{delete} method is
used to define an HTTP \texttt{DELETE} entry point. The arguments to \lstinline[style=javascript]{delete}
include the URL (i.e. \lstinline[style=javascript]{"/users/"}) and path parameter (i.e.
\lstinline[style=javascript]{":userId"}) of the entry point, and the callback function 
that will be executed on incoming requests. The callback function receives request (i.e. \lstinline[style=javascript]{req}) and response (i.e. \lstinline[style=javascript]{res}) 
objects as arguments, reads the \lstinline[style=javascript]{userId} path parameter at 
line 2, and removes the corresponding document from a collection in the database at line 3.
The REST specification, in OpenAPI format \cite{swagger}, corresponding to the example 
entry point of \autoref{example-endpoint} is shown in \autoref{example-inferred-req}. 
Lines 2-4 define the \lstinline[style=javascript]{"paths"} entry that lists valid URL paths, 
where each path contains one entry per valid HTTP method. Lines 5-13 define the
\lstinline[style=javascript]{"parameters"} object that lists the valid parameters for a given
path and HTTP method. Specifically, line 7 defines the name of the parameter, line 8
specifies that it is a path parameter, line 9 specifies that the parameter is required,
line 10 specifies that the expected type of the \lstinline[style=javascript]{"userId"}
parameter is \lstinline[style=javascript]{"string"} and line 11 captures a concrete example
value that was observed while crawling. Directing the crawler to exercise client-side 
code that will trigger server-side endpoints is, however, non-trivial and covered in the 
next section.

\begin{lstlisting}[style=JavaScript, caption={Example endpoint and its callback in Express}, label=example-endpoint, float=bth]
app.delete("/users/:userId", (req, res) => {
  const id = req.params.userId;
  collection.remove({"id": id});
});
\end{lstlisting}

\begin{lstlisting}[style=JavaScript, caption={Automatically generated OpenAPI specification}, label=example-inferred-req, float=thb],
{
  "paths": {
    "/users/{userId}": {
      "delete": {
        "parameters": [
          {
            "name": "userId",
            "in": "path",
            "required": true,
            "type": "string",
            "example": "abc123"
          }
        ]
      }
    }
  }
}
\end{lstlisting}

\begin{lstlisting}[style=JavaScript, caption={Client-side JavaScript code that uses a framework (Angular.js) with customized event handling and registration.}, label=example-client, float=thb],
<html>
<script src='angular.js'></script>
<script>
function foo(){
}
function bar() {
  fetch("users/abc123");
}
</script>
<body>
  <button id='b1' click='foo()'>B1</button>
  <button id='b2' data-ng-click='bar()'>B2</button>
</body>
</html>
  
\end{lstlisting}

\subsection{Prioritised state-aware crawling for REST API inference}

To improve responsiveness, modern web applications often transfer a large portion 
of their logic, including data pre-processing and validation to the client side. 
To prevent errors, frameworks implement checks on the server-side that validate the 
structure and to some extent the content of incoming requests. Fuzzing modern web 
applications thus require to produce requests that get past those initial server-side checks. 
To this end, \textsc{BackREST} infers REST APIs from the requests generated by a
state-aware crawler.
Our crawler extends the one presented in \cite{hassanshahi2020gelato} to 
automatically discover endpoints, parameters and types, and to produce concrete examples, 
as shown in \autoref{example-inferred-req}.

\begin{figure*}
	\centering
	\includegraphics[width=0.7\textwidth]{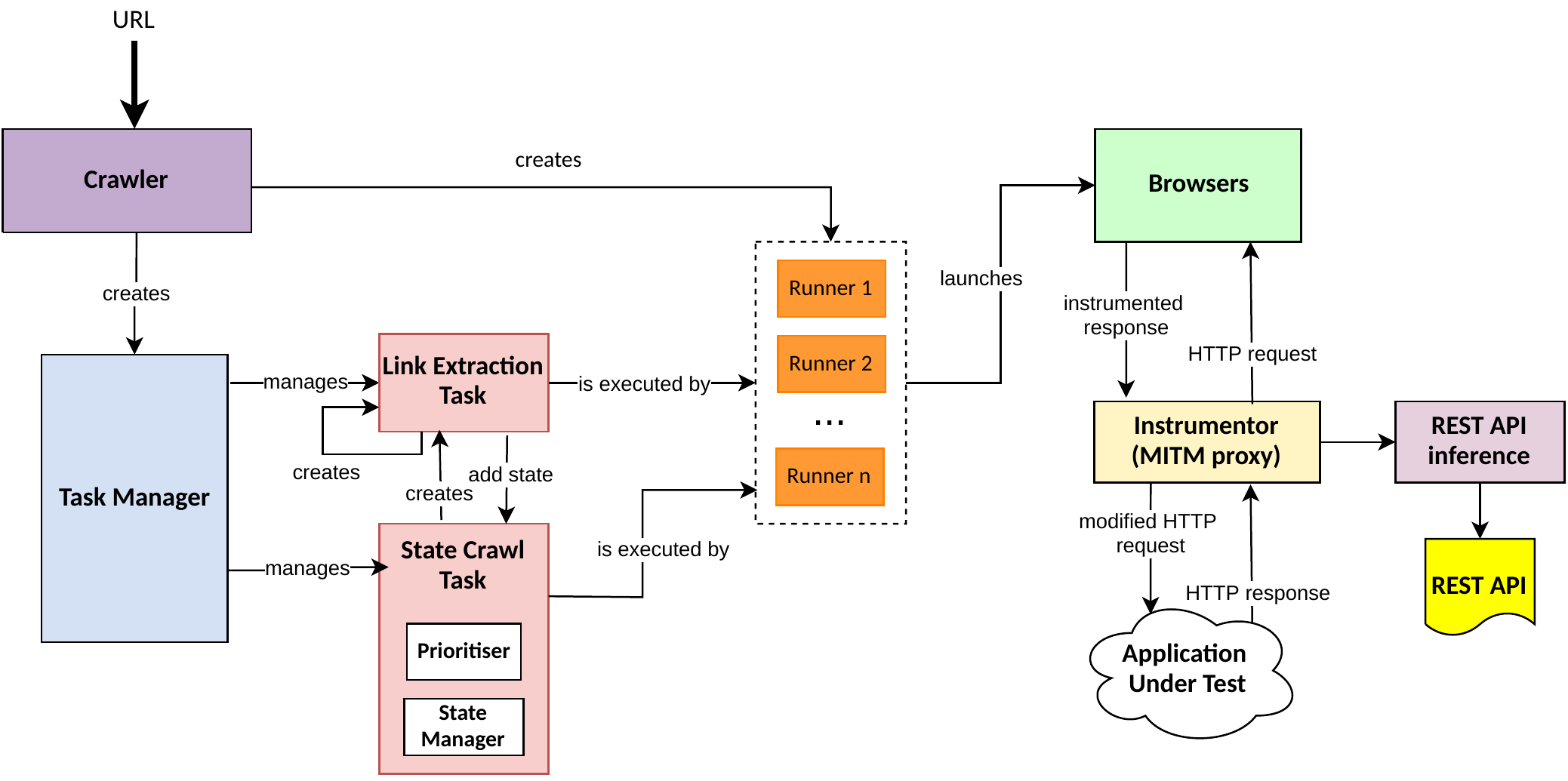}
	\caption{State-aware crawler architecture}
	\label{fig:gelato}
\end{figure*}

\autoref{fig:gelato} shows the architecture of the crawler in \cite{hassanshahi2020gelato}.
It combines prioritized link extraction (or spidering) and state-aware
crawling to support both multi-page and single-page applications.
It performs static link extraction and dynamic state processing
using tasks that are managed in parallel.
Given the URL of a running instance, the crawler's task manager first creates
a link extraction task for the top level url, where
static links are extracted from HTML elements.
Next, new link extraction tasks
are created recursively for newly discovered links.
At the same time, it creates new states for each extracted link and add them
to a priority queue in a state crawl task. The state queue is prioritized either
in a Depth First Search (DFS) or Breadth First Search (BFS) order, depending on
the structure of the application. For instance, if an application
is a traditional multi-page application and most of the endpoints would be triggered
through static links, BFS can be more suitable, whereas a single-page application
with  workflows involving long sequences of client-side interactions, such as
filling input elements and clicking buttons, would benefit more
from a DFS traversal. 

A state in the crawler includes the URL of the loaded page, its DOM tree, static links and valid events.
To avoid revisiting same states, it compares the URLs and their DOM trees
using a given threshold. As described in \cite{hassanshahi2020gelato},
it marks a state as previously visited if there exists a state in the cache that
has the same URL and a DOM tree that is similar enough to the existing state.
To compare the DOM trees, it parses them using the ElementTree XML~\cite{element-tree} 
library and considers two trees to be in the same equivalence class if the number of 
different nodes does not exceed a threshold.

The crawler transitions between states by automatically triggering events
that it extracts from HTML elements. It supports both statically and
dynamically registered events, as well as customized event registration in frameworks.
Determining the priority of events is one of the differentiating factors
between state-aware crawlers. On the one hand, Crawljax~\cite{mesbah2008crawling},
a well-known crawler for AJAX-driven applications randomly selects events from a state.
On the other hand, Artemis~\cite{artemis}, selects events that are more likely to 
increase code coverage.
Feedex~\cite{feedex} instead prioritizes events that trigger user-specified
workflows, such as adding an item. j\"{A}k \cite{jaek} uses dynamic analysis to
create a navigation graph with dynamically generated URLs and
traces that contain runtime information for events. Its crawler then navigates the graph in an attempt to maximise server-side coverage. Similar to j\"{A}k, we also perform dynamic analysis to detect
dynamically registered events that are difficult to detect statically and 
maximise coverage of server-side endpoints. Contrary to j\"{A}k,
however, our crawler also uses a call graph analysis of the JavaScript code to 
compute a distance metric from event handler functions to target functions like \lstinline[style=javascript]{XMLHttpRequest.send()}, 
\lstinline[style=javascript]{HTMLFormElement.submit()}, or \lstinline[style=javascript]{fetch}
and prioritize events that have smaller distances. This feature allows reaching target functions 
that are deeply embedded not only in the application code but also
in the JavaScript frameworks and libraries that it uses. 
This prioritization strategy allows \textsc{BackREST} to maximize coverage of
JavaScript functions that trigger server-side endpoints.
\autoref{example-client} shows a hande-made and very simplified
code-snippet that uses AngularJS for event handling.
While the first button, \codett{B1} is a normal button and uses standard
\codett{click} event registration, the second button, \codett{B2} uses a custom event registration
from the AngularJS framework. Our crawler can successfully correlate the customized \codett{click}
event for button \codett{B2} and prioritize \codett{B2} over \codett{B1} to call
 \codett{fetch("users/abc123")} and trigger the server-side \codett{users/abc123} endpoint.


One of the challenges in dynamic analysis of web applications is performing
authentication and correctly maintaining
the authenticated sessions. Our crawler provides support for a wide variety
of authentication mechanism, including Single Sign On, using a record-and-replay mechanism.
We require the user to record the authentication once and use it to
authenticate the application as many times as required.
To verify whether a session is valid, we ask the user to provide an endpoint and
pattern to look up in the response content once and before the crawling starts.
For instance, for Juice-shop (see Section \ref{sec:evaluation}),
we verify the session by sending
a GET request to the \lstinline[style=javascript]{rest/user/whoami} endpoint and check if \lstinline[style=javascript]{"admin@juice-sh.op"}
is present in its response content periodically to make sure it is logged in.

We intercept the requests triggered by our crawler using
a Man-In-The-Middle (MITM) proxy. Next, we process the recorded HTTP
requests to infer a REST API specification automatically.
Because we use a client-side crawler to trigger the endpoints, the recorded traffic 
contains valid headers and parameter values that are persisted in the REST API
and reused in the fuzzing phase. These seed values can often prove invaluable to 
get past server-side value checks. Our API inference also aggregates concrete values
to infer their types. 
Going back to the example in \autoref{example-endpoint}, by automatically triggering
\lstinline[style=javascript]{delete} requests to \lstinline[style=javascript]{/users/} endpoint with an actual \lstinline[style=javascript]{userId}
string value (e.g. \lstinline[style=javascript]{abc123}), our REST API inference adds \lstinline[style=javascript]{userId} path parameter with
\lstinline[style=javascript]{string} type to the specification based on the observed \lstinline[style=javascript]{userId} values.

\begin{figure*}
	\centering
	\includegraphics[width=0.9\textwidth]{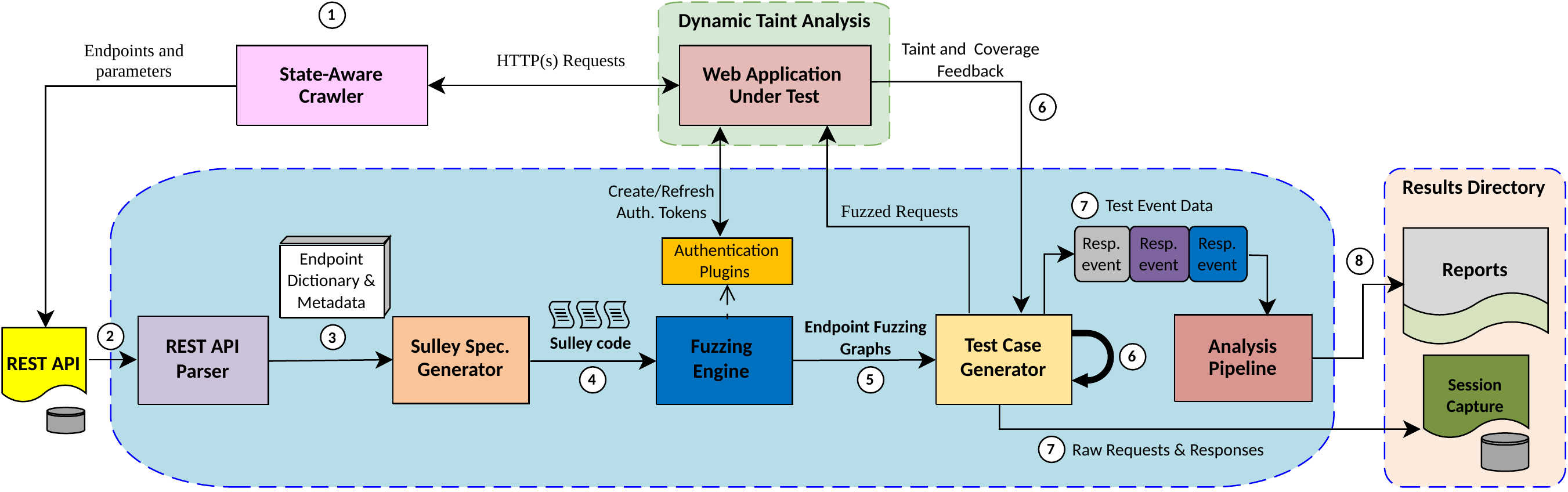}
	\caption{\textsc{BackREST} architecture}
	\label{fig:backrest-arch}
\end{figure*}

\section{Feedback-driven Fuzzing}
\label{sec:feedback-fuzzing}

\textsc{BackREST} builds on top of Sulley \cite{sulley}, a blackbox
fuzzer with built-in networking support, and extends it with support for REST API parsing 
and fuzzing, as well as coverage and taint feedback. Figure~\ref{fig:backrest-arch} shows 
the high-level architecture of \textsc{BackREST}. First, the application under test is crawled
and a REST API is derived \circled{1}. Then, \textsc{BackREST} is invoked on the generated 
REST API file \circled{2}. The API is then parsed \circled{3} and broken down into 
low-level fuzzing code blocks \circled{4}. The type information included in the API 
file is used to select relevant mutation strategies (e.g. string, integer, JWT token, etc.). 
The fuzzing engine is then responsible for evaluating the individual fuzzing code blocks and 
reassembling them into a graph that will yield well-formed HTTP requests \circled{5}. The 
test case generator repeatedly traverses the graph to generate concrete HTTP requests, sends
them to the application under test, and monitors taint and coverage feedback \circled{6}.
The HTTP responses are dispatched to the analysis pipeline, which runs in a separate thread, 
to detect exploits, and are also stored as-is for logging purposes \circled{7}. Indicators of 
exploitation include the response time, the error code and error messages, reflected payloads, 
and taint feedback. Finally, the analysis results are aggregated and reported \circled{8}. To 
simplify our evaluation setup (section \ref{sec:evaluation}), we run \textsc{BackREST} in 
\emph{deterministic} mode, meaning that it always yields the same sequence of fuzzed HTTP 
requests for a given configuration.


\subsection{Coverage Feedback}

Coverage feedback, where the fuzzer uses online coverage information to guide the 
fuzzing session, was made popular by the AFL fuzzer \cite{afl}. Nowadays, 
most greybox fuzzers use coverage to guide their input generation engine towards
producing input that will exercise newly covered code, or branches that will likely lead 
to new code \cite{li2017steelix,lemieux2018fairfuzz,gan2018collafl,petsios2017slowfuzz,lemieux2018perffuzz,bohme2017coverage,bohme2017directed}. 
Kelinci~\cite{kelinci} ported AFL-style greybox fuzzing to Java programs by emulating AFL's 
coverage analysis and using the AFL fuzzing engine as-is. JQF~\cite{jqf} instead combines 
QuickCheck-style testing~\cite{quickcheck} and feedback-directed fuzzing to support inputs 
with arbitrary data structures.
The underlying assumption in AFL and all its derivatives is that targeting code that has 
not been thoroughly exercised increases the likelihood of triggering bugs and vulnerabilities. 
Empirical evidence suggests that this assumption holds true for many codebases. Furthermore,
the simplicity and widespread availability of coverage analysis makes it suitable for 
applications written in a wide range of languages.

Compared to mutation-based greybox fuzzers like AFL, \textsc{BackREST} uses coverage
information differently. Where AFL-like fuzzers use coverage information to \emph{derive}
the next round of input, \textsc{BackREST} uses coverage information to \emph{skip} inputs
in the test plan that would likely exercise well-covered code. From that perspective, \textsc{BackREST} uses coverage information as a performance optimisation. Section \ref{sec:fuzzing-algo} details how \textsc{BackREST} uses coverage information.

\subsection{Taint Feedback}

Taint-directed fuzzing, where the fuzzer uses taint tracking to locate sections of
input that influence values at key program locations (e.g. buffer index, or magic byte checks),
was pioneered by Ganesh et al. with the BuzzFuzz fuzzer \cite{ganesh2009taint}. In recent 
years, many more taint-directed greybox fuzzers that build on the ideas of BuzzFuzz have been
developed \cite{bekrar2012taint, liang2013effective, wang2010taintscope, wang2011checksum, haller2013dowsing, rawat2017vuzzer}.

\textsc{BackREST} uses taint analysis in a different way. With the help of a lightweight 
dynamic taint inference analysis \cite{gauthier2018}, it reports which input reaches security-sensitive 
program locations, and the type of vulnerability that could be triggered at each location. 
Armed with this information, \textsc{BackREST} can prioritise payloads that are more likely 
to trigger potential vulnerabilities. Taint feedback thus enables \textsc{BackREST} to 
\emph{zoom in} payloads that are more likely to trigger vulnerabilities, which improves
performance and detection capabilities. Taint feedback also improves detection capabilities
in cases where exploitation cannot be easily detected in a blackbox manner. Finally, similar
to coverage analysis, the relative simplicity of taint inference analysis makes it easy to port
to a wide range of languages. The next section details how \textsc{BackREST} uses taint feedback 
during fuzzing.

\subsection{\textsc{BackREST} Fuzzing Algorithm}
\label{sec:fuzzing-algo}

\begin{lstlisting}[style=JavaScript, caption={Fuzzable locations for an example endpoint}, label=fuzzable-locations, float=thb],
"/users/(*\bfseries \{userId\}*)": {
  "delete": {
    "parameters": [
      {
        "name": "userId",
        "in": "(*\bfseries path*)",
        "required": (*\bfseries true*),
        "type": "(*\bfseries string*)",
        "example": "abc123"
      }
    ]
  }
}
\end{lstlisting}

\begin{algorithm}
	\SetAlgoLined
	\KwIn{web app. $\mathcal{W}$, REST model $\mathcal{R}$, threshold $\mathcal{T}$, payload dictionary. $\mathcal{D}$}
	\KwOut{vulnerability report $\mathcal{R}$}
	$\mathcal{R} \leftarrow \emptyset$\;
	$\mathcal{P} \leftarrow \textsc{BuildTestPlan}(\mathcal{R})$\;
	$\mathcal{W'} \leftarrow \textsc{coverageInstrument}(\mathcal{W})$\;
	$\mathcal{W''} \leftarrow \textsc{taintInstrument}(\mathcal{W'})$\;
	$totalCov \leftarrow 0$\;
	\ForEach{$endpoint$ in $\mathcal{P}$}{
		\ForEach{$location$ in $\mathcal{P}[endpoint]$} {
			$types \leftarrow \mathcal{D}.keys()$\;
			taint:\;
			\ForEach{$type$ in $types$}{
				coverage:\;
				$count \leftarrow 0$\;
				\ForEach{$p$ in $\mathcal{D}[type]$}{
					$(resp, currCov, taintCat) \leftarrow \textsc{fuzz}(endpoint,location,payload,\mathcal{W''})$\;
					$count \leftarrow count + 1$\;
					\If{$currCov > totalCov$}{
						$count \leftarrow 0$\;
					}
					$vuln \leftarrow \textsc{DetectVulnerability}(resp)$\;
					\If{$vuln \neq \emptyset$}{
						\Print $vuln$\;
					}
					$totalCov \leftarrow currCov$\;
					\If{$taintCat \neq \emptyset$}{
						$types \leftarrow (taintType)$\;
						\If{$taintType \neq types$}{
							\Break taint\;
						}
						$count \leftarrow 0$\;
					}
					\If{$count > \mathcal{T}$}{
						\Break coverage\;
					}
				}	
			}
		}
	}	
	\Return{$\mathcal{V}$}
	\caption{REST-based feedback-driven fuzzing}
	\label{alg:guided_fuzzing}
\end{algorithm}

Algorithm~\ref{alg:guided_fuzzing} shows the \textsc{BackREST} fuzzing algorithm. 
It first builds a test plan, based on the REST model received as input (line 2). The 
test plan breaks the REST model into a set of \emph{endpoints}, lists ``fuzzable'' 
\emph{locations} in each endpoint, and establishes a mutation schedule that specifies 
the values that are going to be injected at each location. Values are either cloned from
the \lstinline[style=javascript]{example} fields, derived using mutations (omitted from 
\autoref{alg:guided_fuzzing} for readability), or drawn 
from a pre-defined dictionary of payloads where vulnerability types map to a set of 
payloads. For example, the SQLi payload set contains strings like: \texttt{' OR '1'='1' --}, 
while the buffer overflow set contains very large strings and numbers. 

\autoref{fuzzable-locations} shows an example of an endpoint definition with fuzzable locations in bold. From top to
bottom, fuzzable locations include: the value of the \texttt{userId} parameter, where the 
parameter will be injected (e.g. path or request body), whether the parameter is required 
or optional, and the type of the parameter. Other locations are left untouched to preserve
the core structure of the model and increase the likelihood of the request getting past 
shallow syntactic and semantic checks.

Once the test plan is built, \textsc{BackREST} instruments the application for coverage 
and taint inference (lines 3-4 of \autoref{alg:guided_fuzzing}). Then, it starts iterating over the endpoints in the test 
plan (line 6). The fuzzer further sends a request for each \texttt{(endpoint, location, 
payload)} combination, collects the response, coverage and taint report, and increases its 
request counter by one (lines 13-15). If the request covers new 
code, the request counter is reset to zero, allowing the fuzzer to spend more time
fuzzing that particular endpoint, location, and vulnerability type (lines 16-18). The 
vulnerability detector then inspects the response, searching for indicators of exploitation, 
and logs potential vulnerabilities (lines 19-22). 

Because blackbox vulnerability detectors inspect the response only, 
they might miss cases where an input reached a security-sensitive sink, without producing an observable 
side-effect. For example, a command injection 
vulnerability can be detected in a blackbox fashion
only when an input triggers an observable side-effect, such as printing a 
fuzzer-controlled string, or making the application sleep for a certain amount of time. 
With taint feedback, however, the fuzzer is informed about: 1. whether parts of the input 
reached a sink, and 2. the vulnerability type associated with the sink. When the
fuzzer is informed that an input reached a sink, it immediately \emph{jumps} to the
vulnerability type that matches that of the sink and starts sending payloads of that type
only (lines 24-30). The idea behind this heuristic is that payloads that match
the sink type have a higher chance of triggering observable side-effects to help 
confirm a potential vulnerability. Targeting specific vulnerability types further
minimises the number of inputs required to trigger a vulnerability. Finally, if a given 
endpoint and location pair have been fuzzed for more than $\mathcal{T}$ requests (10 in
our setup) without increasing coverage, the fuzzer \emph{jumps} 
to the next vulnerability type (lines 31-33). The idea behind this heuristic, which we 
validated on our benchmarks (see \autoref{sec:feedback_eval}), is that the likelihood 
of covering new code by fuzzing a given endpoint decreases with the number of requests.  
Promptly switching to payloads from a different vulnerability type reduces the total 
number of requests, thereby improving the overall performance.

\section{Implementation}
\label{sec:implementation}
\textsc{BackREST} brings together and builds on top of several existing components.
The REST API inference component uses the state-aware crawler from \cite{hassanshahi2020gelato}
with an intercepting proxy \cite{mitmproxy} to generate and capture traffic dynamically.
The REST API parser is derived from PySwagger \cite{pyswagger}
and the fuzzing infrastructure extends Sulley \cite{sulley}, with REST API processing support. 
For coverage instrumentation, we use the Istanbul library \cite{istanbul} and have added 
a middleware in benchmark applications to read the coverage object after each request, 
and inject a custom header to communicate coverage results back to the fuzzer. For taint 
feedback, \textsc{BackREST} implements the Node.js taint analysis from \cite{gauthier2018}
and extends our custom middleware to also communicate taint results back to the fuzzer. 
The taint analysis is itself built on top of the NodeProf.js instrumentation framework 
\cite{sun2018efficient} that runs on the GraalVM \cite{wurthinger2013one} runtime.

\section{Evaluation}
\label{sec:evaluation}


\begin{table}
	\centering
	\caption{Benchmark applications}
	\begin{tabular}{llrrr}
		\toprule
		Application & Description & Version & LOC & Files \\
		\midrule
		Nodegoat & Educational & 1.3.0 & 970\,450 & 12\,180 \\
		Keystone & CMS & 4.0.0 & 1\,393\,144 & 13\,891 \\
		Apostrophe & CMS & 2.0.0 & 774\,203 & 5\,701 \\
		Juice-shop & Educational & 8.3.0 & 725\,101 & 7\,449\\
		Mongo-express & DB manager & 0.51.0 & 646\,403 & 7\,378 \\
		\bottomrule
	\end{tabular}
	\label{tab:benchmarks}
\end{table}

In this section, we first evaluate how coverage and taint feedback increase the coverage, performance, 
or number of vulnerabilities detected. The benchmark applications used for evaluation are listed in 
\autoref{tab:benchmarks}. All experiments were run on a machine with 8 Intel Xeon E5-2690 
2.60GHz processors and 32GB memory. Then, we compare \textsc{BackREST} to three state-of-the-art 
web fuzzers. Finally, we present and explain the 0-days that \textsc{BackREST} detected. 

\begin{figure*}
	\centering
	\subfigure[Nodegoat coverage]{
		\includegraphics[width=0.47\columnwidth]{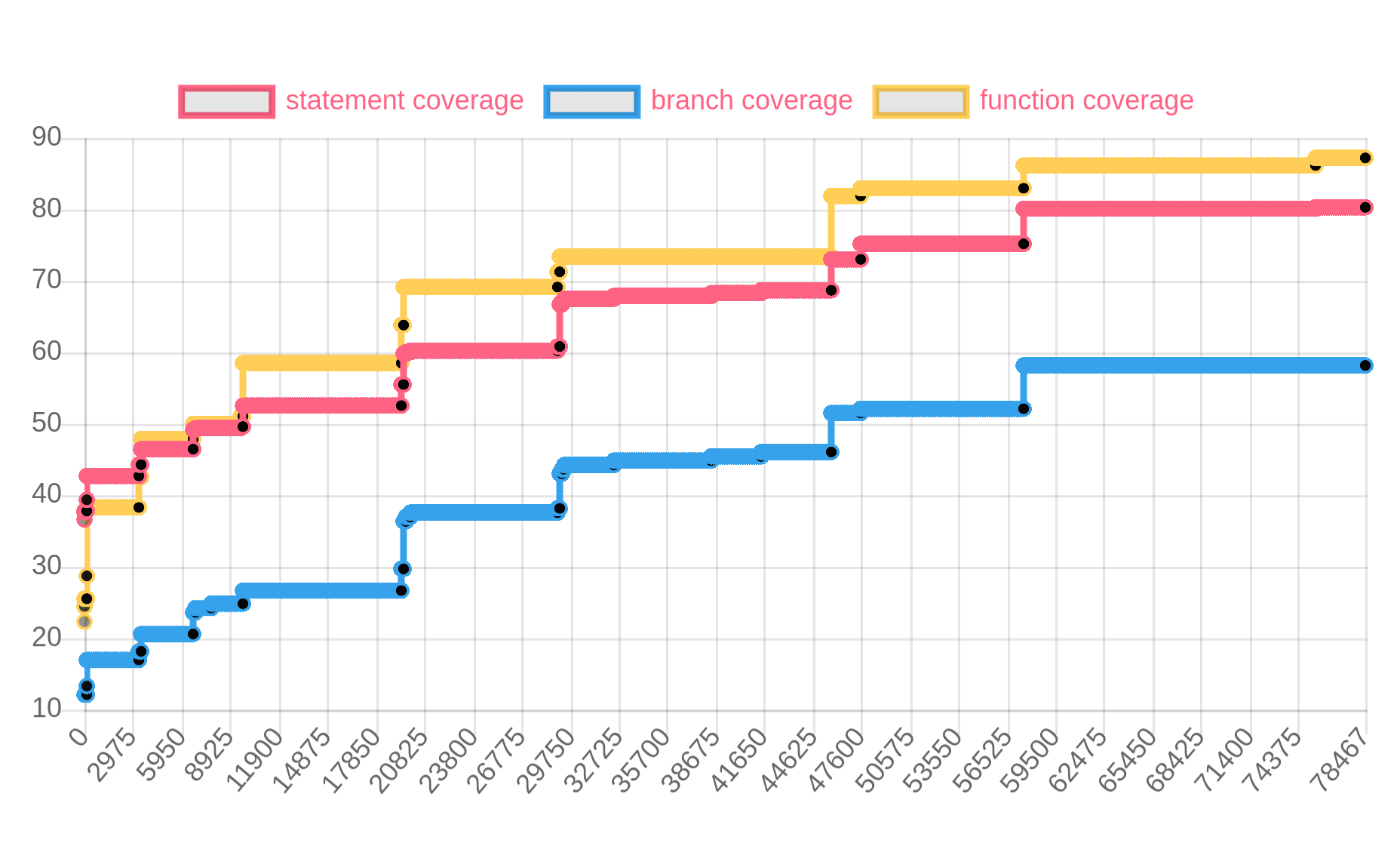}
		\label{fig:nodegoat_cov}
	}
	\subfigure[Keystone coverage]{
		\includegraphics[width=0.47\columnwidth]{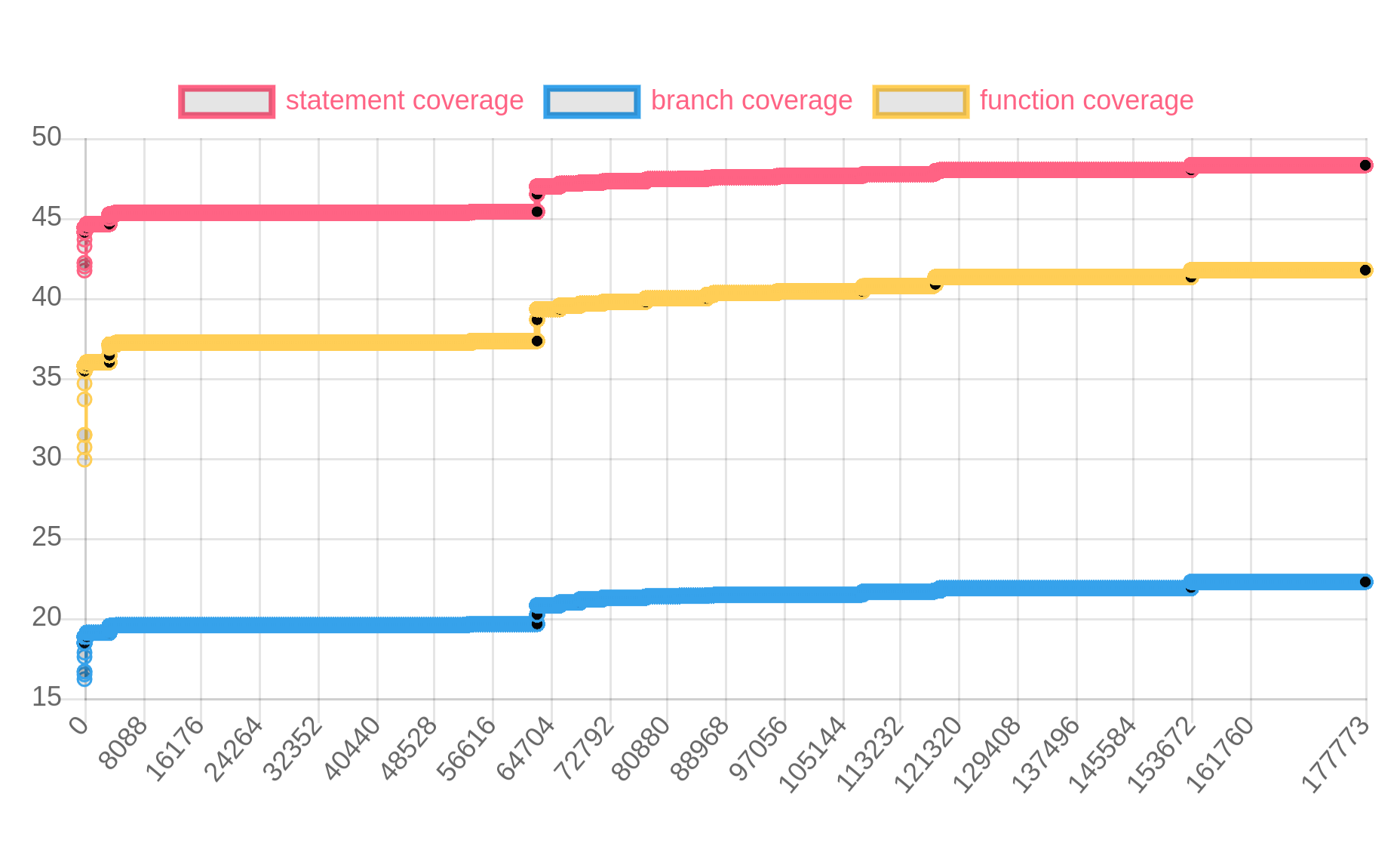}
		\label{fig:keystone_cov}
	}
	\subfigure[Juice-shop coverage]{
		\includegraphics[width=0.47\columnwidth]{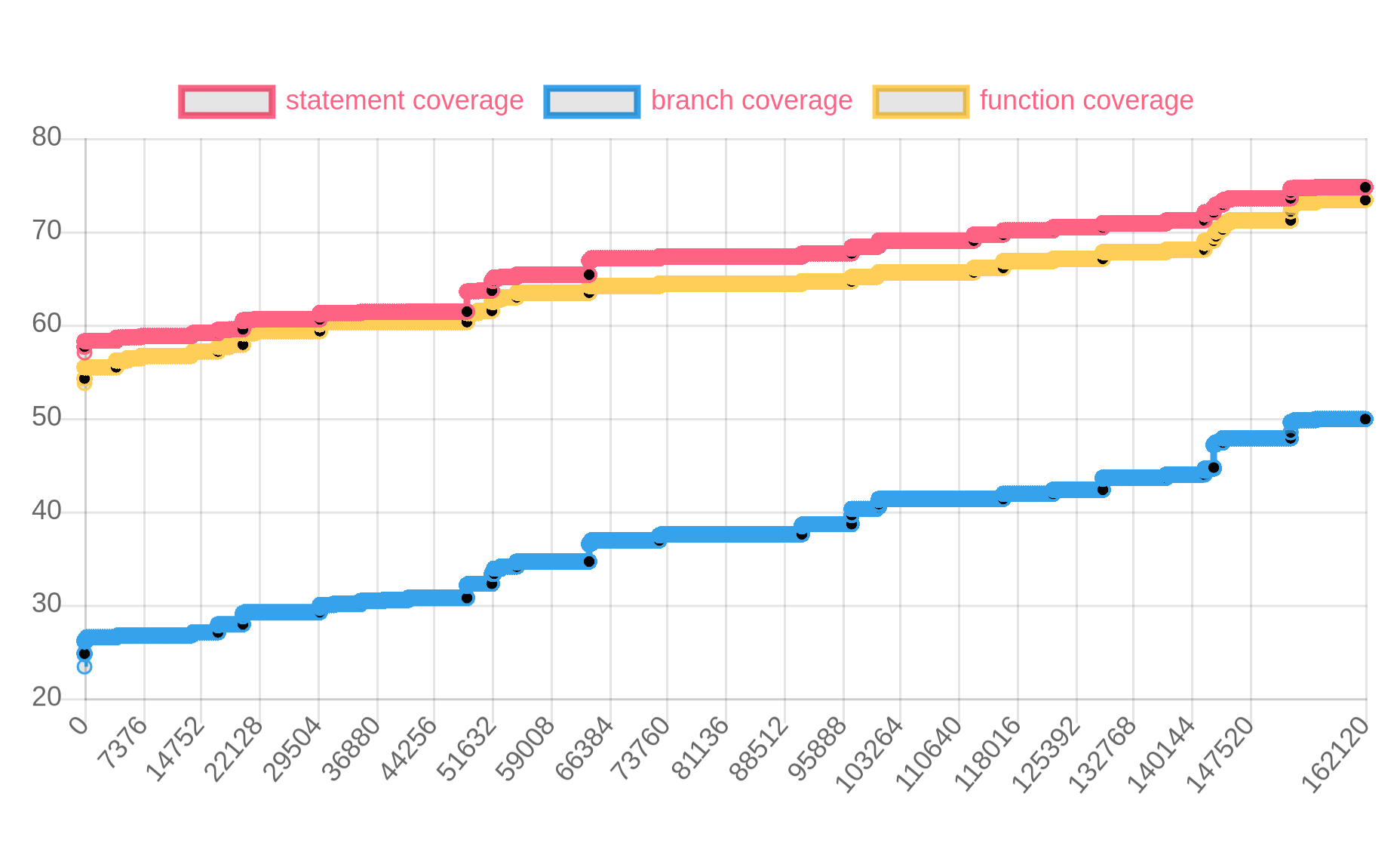}
		\label{fig:juiceshop_cov}
	}
	\subfigure[Mongo-express coverage]{
		\includegraphics[width=0.47\columnwidth]{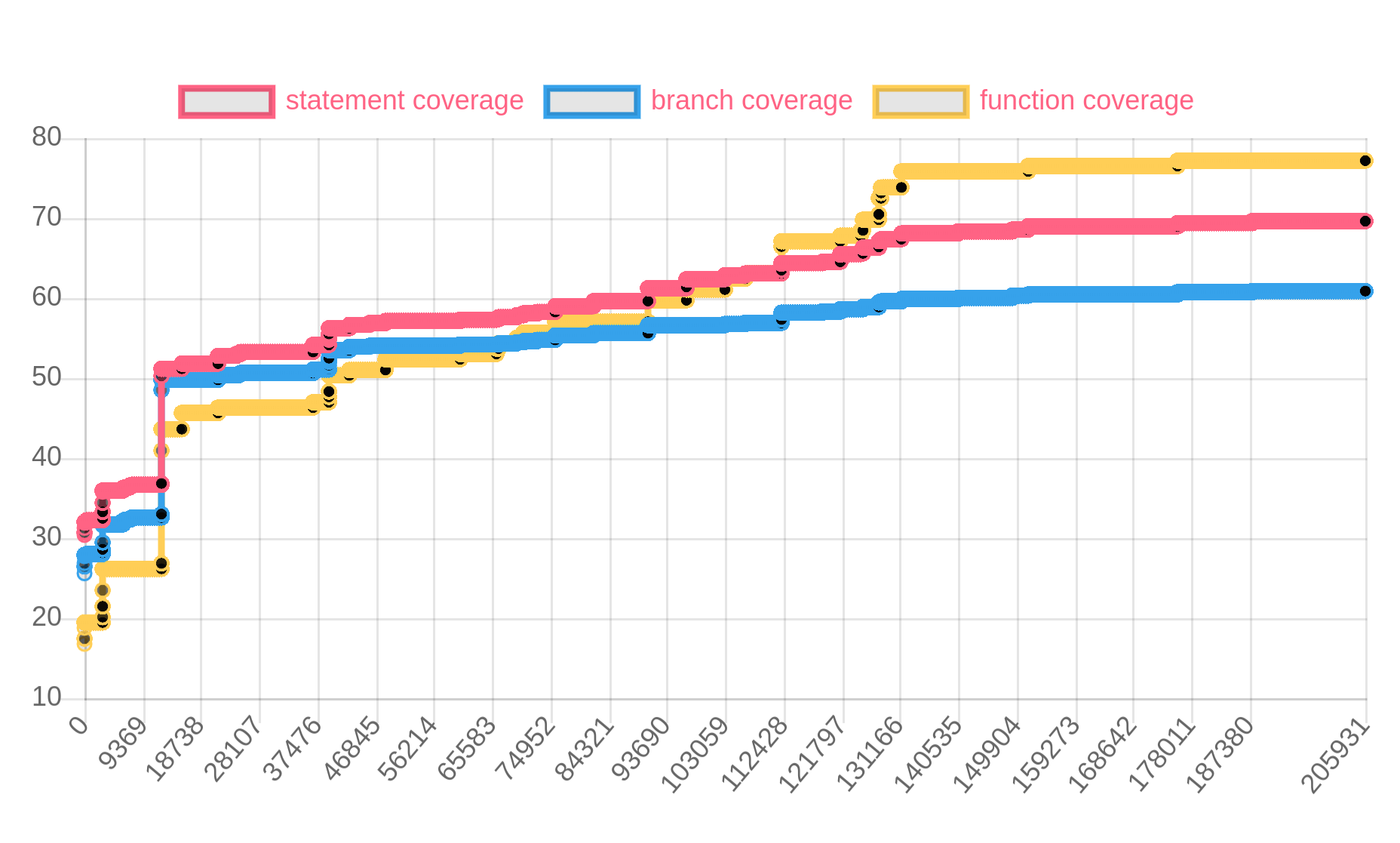}
		\label{fig:mongo_cov}
	}
	\caption{Cumulative statement, branch, and function coverage (y axis) in function 
		of the number of requests (x axis) for Nodegoat \subref{fig:nodegoat_cov},
		Keystone \subref{fig:keystone_cov}, Juice-shop \subref{fig:juiceshop_cov}, and
		Mongo-express \subref{fig:mongo_cov} with the baseline blackbox fuzzer}
	\label{fig:coverage}
\end{figure*}

\subsection{Feedback-driven fuzzing}
\label{sec:feedback_eval}

\begin{table*}
	\centering
	\caption{Impact of the coverage feedback loop on runtime and total coverage}
	\begin{tabular}{lrrrrrrrr}
		\toprule
		\multirow{2}{*}{Benchmark} & \multicolumn{3}{c}{Coverage (\%)} & \multicolumn{5}{c}{Time (hh:mm:ss)}\\
		\cmidrule(lr{.75em}){2-4} \cmidrule(lr{.75em}){5-9} 
		& \multicolumn{1}{c}{B} & \multicolumn{1}{c}{C} & \multicolumn{1}{c}{CT} & \multicolumn{1}{c}{B} & \multicolumn{2}{c}{C} & \multicolumn{2}{c}{CT} \\
		\midrule
		\begin{filecontents*}{data/coverage_feedback.csv}
dummy,Benchmark,timeBaseline,timeFeedback,dummy,timeTaint,dummy,coverageBaseline,coverageFeedback,coverageTaint
dummy,Nodegoat,0:42:39,0:06:07, (7.0$\times$),00:05:44, (7.4$\times$),80.31,78.54,75.59
dummy,Keystone,5:46:29,0:49:25, (7.0$\times$),0:13:23, (25.9$\times$),48.31,48.05,45.43
dummy,Apostrophe,---,11:11:42, --,6:17:34, --,---,48.40,45.52
dummy,Juice-Shop,12:48:15,1:10:31, (10.9$\times$),1:08:26, (11.2$\times$),74.73,76.34,75.85
dummy,Mongo-express,2:21:49,0:16:07, (8.8$\times$),0:11:07, (12.8$\times$),69.62,69.57,66.59
\end{filecontents*}
\csvreader[late after line=\\]{data/coverage_feedback.csv}{}{\csvcolii & \csvcolviii & \csvcolix & \csvcolx & \csvcoliii & \csvcoliv & \csvcolv & \csvcolvi & \csvcolvii}
		\bottomrule
	\end{tabular}
	\label{tab:coverage_feedback}
\end{table*}

\autoref{tab:coverage_feedback} compares the total coverage and runtime achieved by
enabling the coverage feedback loop only (column C) and combined with taint feedback 
(column CT) against the baseline blackbox fuzzer (column B). 
\autoref{tab:coverage_feedback} also lists speedups for coverage and taint feedback
loops compared to baseline. Coverage-wise, enabling the coverage feedback loop, 
which skips payloads of a given type after $\mathcal{T}$ requests that did not
increase coverage, achieves approximately the same coverage (i.e. $\pm$ 2\%) 
in a much faster way (i.e. speedup between $7.0\times$ and $10.9\times$). 
The slight variations in coverage can be explained by many different factors, such as
the number of dropped requests, differences in scheduling and number of asynchronous 
computations, and differences in the application internal state. Indeed, the process
of fuzzing puts the application under such a heavy load that exceptional behaviours 
become more common. Adding taint feedback on top of coverage feedback further 
decreases runtime, with speedups between $7.4\times$ and $25.9\times$. The slightly 
lower coverage can be explained by the fact that taint feedback forces the fuzzer to
skip entire payload types, resulting in lower input diversity and slightly lower 
total coverage. Finally, the size of the REST model for Apostrophe and the load that
resulted from using the baseline fuzzer rendered the application unresponsive, and we
killed the fuzzing session after 72 hours. Enabling taint feedback for Apostrophe
almost halved the runtime compared to coverage feedback alone.

\autoref{fig:coverage} shows the cumulative coverage achieved by the baseline blackbox
fuzzer on all applications but Apostrophe. For all applications, cumulative coverage 
evolves in a step-wise fashion (e.g. marked increases, followed by plateaus) where 
steps correspond to the fuzzer switching to a new endpoint. The plateaus that 
follow correspond to the fuzzer looping through its payload dictionary. 
These results support our coverage feedback heuristic, which is based on the 
assumption that the likelihood of covering new code by fuzzing a given endpoint 
decreases with the number of requests.

\subsection{A note on server-side state modelling}
\label{sec:stateful}
Many studies have shown that state-aware crawling of the client-side yields better coverage~\cite{mesbah2008crawling, artemis, feedex, jaek, doupe2012enemy}, and our
crawler is no exception. Very little is known, however, about the impact of state-aware fuzzing
of the server-side. To our knowledge, RESTler \cite{atlidakis2019restler} is the first study to 
investigate stateful fuzzing of web services. While the authors have found a positive correlation
between stateful fuzzing and increases in coverage, we have not observed a similar effect on our
benchmark applications. Similar to RESTler, we attempted to model the state of our benchmark 
applications by inferring dependencies between endpoints. Specifically, we used the approach 
from \cite{beschastnikh2011leveraging} to infer endpoint dependencies from crawling logs and
then constrained the fuzzing schedule of \textsc{BackREST} to honour them. This did not improve
coverage for all but the Mongo-express application (data not shown). In this particular case,
manual inspection revealed that the inferred dependencies were quite intuitive (e.g. insert a
document before deleting it) and easily configured. 

\subsection{Vulnerability detection}

\begin{table*}
	\centering
		\caption{Impact of the coverage feedback loop on bug reports}
		\begin{tabular}{lcccccccccccc}
			\toprule
			\multirow{2}{*}{Benchmark} & \multicolumn{3}{c}{(No)SQLi} & \multicolumn{3}{c}{Cmd injection} & \multicolumn{3}{c}{XSS} & \multicolumn{3}{c}{DoS} \\
			\cmidrule(lr{.5em}){2-4} \cmidrule(lr{.5em}){5-7} \cmidrule(lr{.5em}){8-10} \cmidrule(lr{.5em}){11-13}
			& \multicolumn{1}{c}{B} & \multicolumn{1}{c}{C} & \multicolumn{1}{c}{CT} & \multicolumn{1}{c}{B} & \multicolumn{1}{c}{C} & \multicolumn{1}{c}{CT} & \multicolumn{1}{c}{B} & \multicolumn{1}{c}{C} & \multicolumn{1}{c}{CT} & \multicolumn{1}{c}{B} & \multicolumn{1}{c}{C} & \multicolumn{1}{c}{CT} \\
			\midrule
			\begin{filecontents*}{data/coverage_feedback_bugs.csv}
dummy,Benchmark,SQLiB,SQLIC,SQLIT,RCEB,RCEC,RCET,XSSB,XSSC,XSST,DoSB,DoSC,DoST
dummy,Nodegoat,0,0,3,0,0,3,5,5,5,0,0,0
dummy,Keystone,0,0,0,0,0,0,1,1,0,0,0,0
dummy,Apostrophe,0,0,0,0,0,0,1,1,1,2,1,1
dummy,Juice-Shop,1,1,2,0,0,1,4,1,1,1,0,0
dummy,Mongo-express,0,0,5,0,0,2,0,0,0,3,3,3
\end{filecontents*}
\csvreader[late after line=\\]{data/coverage_feedback_bugs.csv}{}{\csvcolii & \csvcoliii & \csvcoliv & \csvcolv & \csvcolvi & \csvcolvii & \csvcolviii & \csvcolix & \csvcolx & \csvcolxi & \csvcolxii & \csvcolxiii & \csvcolxiv}
			\midrule
			Total & 1 & 1 & 10 & 0 & 0 & 6 & 11 & 8 & 7 & 6 & 4 & 4 \\
			\bottomrule
		\end{tabular}
		\label{tab:coverage_feedback_bug}
\end{table*}

\autoref{tab:coverage_feedback_bug} shows the number of unique true positive bug reports 
with the baseline fuzzer (column B), with coverage feedback (column C), and further
adding taint feedback (column CT). We manually reviewed all reported vulnerabilities
and identified the root causes. \autoref{tab:coverage_feedback_bug} does not list false 
positives for space and readability reasons. The only false positives were an XSS in 
Nodegoat that was reported by all three variations, and an SQLi in Keystone that was 
reported with taint feedback only. Also note that \autoref{tab:coverage_feedback_bug} 
lists three types of vulnerabilities (SQLi, command injection, and XSS) and one type 
of attack (DoS). We opted to list DoS for readability reasons. Indeed, the root 
causes of DoS are highly diverse (out-of-memory, infinite loops, uncaught 
exception, etc.) making it difficult to list them all. From our experience, 
the most prominent root cause for DoS in Node.js are uncaught exceptions. 
Indeed, contrary to many web servers, the Node.js front-end that listens to incoming 
requests is single-threaded. Crashing the front-end thread with an uncaught exception 
thus crashes the entire Node.js process \cite{ojamaa2012assessing}.

Interestingly, enabling coverage feedback has no impact on the detection of SQLi and 
command injection vulnerabilities, suggesting that this optimisation could be enabled at no cost.
Enhancing the fuzzer with taint feedback, however, consistently detects as many or
more SQLi and command injection vulnerabilities. This is explained by the fact that taint inference
does not rely on client-observable 
side-effects of a payload to detect vulnerabilities. This is especially obvious 
for command injection vulnerabilities, which are detected with taint feedback only,
for which observable side-effects are often hard to correlate to the root cause 
(e.g. slowdowns, internal server errors), compared to cross-site scripting, for
example. Because the most critical injection flaws sit on the server-side of web 
applications, and are, by nature, harder to detect at the client side, the taint 
inference in \textsc{BackREST} gives it a tremendous edge over blackbox fuzzers. 
\autoref{tab:coverage_feedback_bug} also shows, however, that some cross-site 
scripting and denial-of-service vulnerabilities are missed when coverage and taint feedback are enabled. First, all missed XSS are stored XSS. 
Indeed, through sheer brute force, the baseline fuzzer manages to send very specific 
payloads that exploit stored XSS vulnerabilities and trigger side-effects that can be observed at the client
(e.g. reflecting the payload in another page), while coverage and taint feedback loops 
caused these specific payloads to be skipped. To reliably detect stored XSS, 
taint analysis would need to track taint flows through storage, which implies 
shadowing every storage device (i.e. databases and file system) to store and 
propagate taint. This feature is beyond greybox fuzzing and is known to be tricky 
to implement, and costly from both time and memory perspectives \cite{kieyzun2009automatic}. 
Second, the missed denial-of-service vulnerabilities are due to: 1. a slow memory 
leak that requires several thousand requests to manifest in Apostrophe, and 2. 
a specific SQLite input that happened to be skipped with coverage and taint feedback.

\paragraph{Evaluating false negatives} In the context of a fuzzing session, false
negatives are those vulnerabilities that are in the scope of a fuzzer, but that are 
missed. Accounting for false negatives requires an application with known 
vulnerabilities. The OWASP Nodegoat and Juice-Shop projects are deliberately vulnerable applications with seeded vulnerabilities. Both projects were built for
educational purposes, and both have official 
solutions available, making it possible to evaluate false negatives. The solutions,
however, list vulnerabilities from a penetration tester perspective; they list attack
payloads, together with the error messages or screens they should lead to. For this 
reason, correlating the official solutions to \textsc{BackREST} reports is not trivial. 
For example, the Juice-Shop solution reports several possible different SQL 
and NoSQL injection attacks. From a fuzzing perspective, however, all these attacks 
share the same two root causes: calling specific MongoDB and Sequelize query methods 
with unsanitised inputs. In other words, while the official solutions report 
different exploits, \textsc{BackREST} groups them all under the same two 
vulnerability reports. For this study, we manually correlated all the SQLi, command injection, 
XSS, and DoS exploits in official solutions to vulnerabilities in the applications 
and found that \textsc{BackREST} reports them all, achieving a recall of 100\% for 
Nodegoat and Juice-Shop.

\begin{figure*}
	\centering
	\begin{tikzpicture}  
	\begin{axis}
	[
	axis lines=left,
	ybar,
	bar width=6pt,
	enlargelimits=0.1,
	width=14cm,height=60mm,
	legend style={at={(0.45,-0.5)}, anchor=north,legend columns=-1},     
	ylabel={Coverage (\%)}, 
	symbolic x coords={Nodegoat, Keystone, Apostrophe, Juice-Shop, Mongo-express},  
	xtick=data,  
	xticklabel style={rotate=45},
	nodes near coords,  
	every node near coord/.append style={rotate=90,anchor=west},
	cycle list/Dark2,
	every axis plot/.append style={fill}
	]
	
	\addplot coordinates {(Nodegoat, 75.59)(Keystone, 45.43)(Apostrophe, 45.52)(Juice-Shop, 75.85)(Mongo-express, 66.59)};
	\addplot coordinates {(Nodegoat, 89.11)(Keystone, 48.94)(Apostrophe, 43.11)(Juice-Shop, 63.36)(Mongo-express, 67.56)};		
	\addplot coordinates {(Nodegoat, 75.10)(Keystone, 46.83)(Apostrophe, 40.80)(Juice-Shop, 62.17)(Mongo-express, 66.42)};
	\addplot coordinates {(Nodegoat, 69.96)(Keystone, 46.15)(Apostrophe, 40.41)(Juice-Shop, 63.17)(Mongo-express, 65.37)};
	\legend{\textsc{BackREST}, Zap, Arachni, w3af}
	
	\end{axis}  
	\end{tikzpicture}
	\caption{Coverage achieved by different fuzzers.}
	\label{fig:coverage-comparison}
\end{figure*}
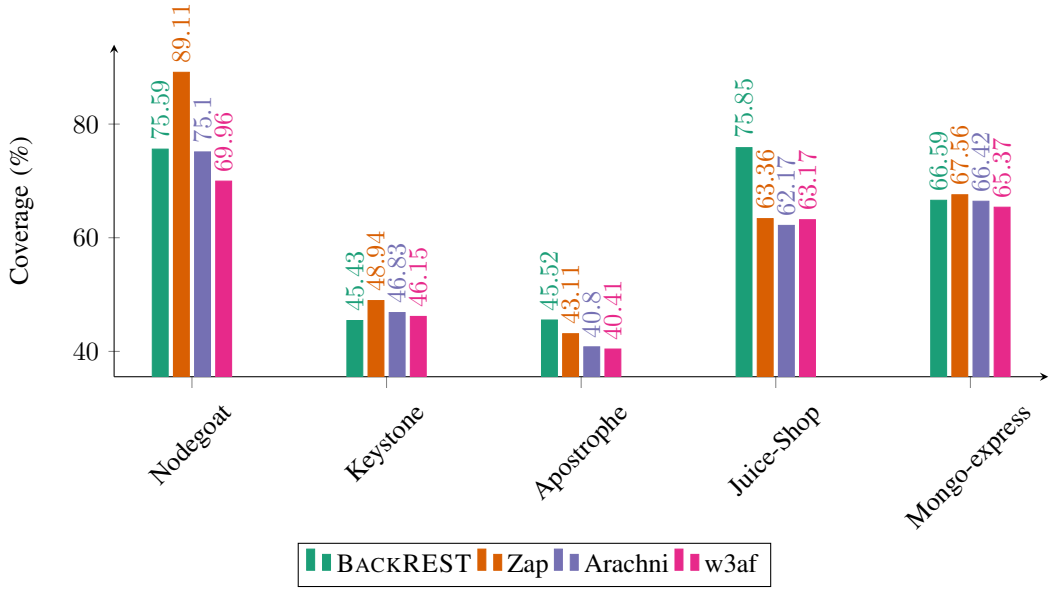

\begin{table*}
	\centering
	\caption{Number of vulnerabilities reported per fuzzer}
	\begin{tabular}{lcccccccccccccccc}
		\toprule
		\multirow{2}{*}{Benchmark} & \multicolumn{4}{c}{(No)SQLi} & \multicolumn{4}{c}{Command injection} & \multicolumn{4}{c}{XSS} & \multicolumn{4}{c}{DoS} \\
		\cmidrule(lr{.5em}){2-5} \cmidrule(lr{.5em}){6-9} \cmidrule(lr{.5em}){10-13} \cmidrule(lr{.5em}){14-17}
		& \multicolumn{1}{c}{BR} & \multicolumn{1}{c}{Zap} & \multicolumn{1}{c}{Arch} & \multicolumn{1}{c}{w3af}
		& \multicolumn{1}{c}{BR} & \multicolumn{1}{c}{Zap} & \multicolumn{1}{c}{Arch} & \multicolumn{1}{c}{w3af}
		& \multicolumn{1}{c}{BR} & \multicolumn{1}{c}{Zap} & \multicolumn{1}{c}{Arch} & \multicolumn{1}{c}{w3af}
		& \multicolumn{1}{c}{BR} & \multicolumn{1}{c}{Zap} & \multicolumn{1}{c}{Arch} & \multicolumn{1}{c}{w3af} \\
		\midrule
		\begin{filecontents*}{data/fuzzer_bugs.csv}
dummy,Benchmark,SQLIBR,SQLIZ,SQLIA,SQLIW,RCEBR,RCEZ,RCEA,RCEW,XSSBR,XSSZ,XSSA,XSSW,DOSBR,DOSZ,DOSA,DOSW
dummy,Nodegoat,3,3,0,2,3,0,0,3,5,4,2,3,0,0,0,0
dummy,Keystone,0,0,0,0,0,0,0,0,0,0,0,0,0,0,0,0
dummy,Apostrophe,0,0,0,0,0,0,0,0,1,0,0,0,1,0,0,0
dummy,Juice-Shop,2,1,1,1,1,0,0,0,1,1,0,0,0,0,0,0
dummy,Mongo-express,5,0,0,0,2,0,0,0,0,0,0,0,3,4,3,3
\end{filecontents*}
\csvreader[late after line=\\]{data/fuzzer_bugs.csv}{}{\csvcolii & \csvcoliii & \csvcoliv & \csvcolv & \csvcolvi & \csvcolvii & \csvcolviii & \csvcolix & \csvcolx & \csvcolxi & \csvcolxii & \csvcolxiii & \csvcolxiv & \csvcolxv & \csvcolxvi & \csvcolxvii & \csvcolxviii}
		\midrule
		Total & \textbf{10} & 4 & 1 & 3 & \textbf{6} & 0 & 0 & 3 & \textbf{7} & 5 & 2 & 3 & \textbf{4} & \textbf{4} & 3 & 3\\
		\bottomrule
	\end{tabular}
	\label{tab:fuzzer_bugs}
\end{table*}

\subsection{Comparison with state-of-the-art}

In this section, we compare \textsc{BackREST} against the arachni \cite{arachni}, w3af
\cite{w3af} and OWASP Zap \cite{zap} blackbox web application fuzzers. To minimise bias and ensure
a fair evaluation, all three fuzzers were evaluated by an independent party who did not
contribute and did not have access to \textsc{BackREST}. All fuzzers were configured
to scan for (No)SQLi, command injection, XSS, and DoS vulnerabilities. Significant care was also
taken to configure all the fuzzers to authenticate into the applications and not log 
themselves out during a scan. Finally, after we discovered that the crawlers in arachni 
and w3af are fairly limited when it comes to navigating single-page web applications
that heavily rely on client-side JavaScript, we evaluated these fuzzers with seed URLs 
from a Zap crawling session. Zap internally uses the Crawljax \cite{mesbah2008crawling} 
crawler that is better suited to 
navigate modern JavaScript-heavy applications. 

Figure \ref{fig:coverage-comparison} shows the coverage that was achieved by the
different fuzzers on the benchmark applications. \textsc{BackREST} 
consistently achieves comparable coverage to other fuzzers. Table \ref{tab:fuzzer_bugs} 
compares the number of bugs found by each fuzzer. For \textsc{BackREST}, we report the 
bugs found with coverage and taint feedback only. Apart from denial-of-service, 
\textsc{BackREST} consistently detects more vulnerabilities than other fuzzers, which we 
mostly attribute to the taint feedback loop. Indeed, while blackbox fuzzers can only observe
the \emph{side-effects} of their attacks through error codes and client-side inspection,
\textsc{BackREST} can determine with high precision if a payload reached a sensitive
sink and report vulnerabilities that would otherwise be difficult to detect in a purely 
blackbox fashion. Furthermore, we confirmed through manual inspection that apart from
DoS, \textsc{BackREST} always reports a strict superset of the vulnerabilities reported
by the other fuzzers. Deeper inspection revealed that the additional DoS found by Zap
in Mongo-Express was due to a missing URL-encoded null byte payload (\texttt{\%00}) in 
our payload dictionary. 

It is very difficult to compare the performance of different web fuzzers, given the number 
of tunable parameters that each of them offers. For this reason, we focused our efforts on 
configuring them to maximise their detection power and did let them run until completion.
Qualitatively, none of the fuzzer emerged as significantly faster or slower than the others.
The runtime of \textsc{BackREST}, as reported in \autoref{tab:coverage_feedback}, is directly
proportional to the size of the REST API, which explains the longer runtime on Apostrophe.

\begin{table*}
	\centering
	\begin{threeparttable}
		\caption{0-day vulnerabilities}
		\begin{tabular}{lllcll}
			\toprule
			Codebase & Vulnerability & Found by & Taint only & Severity & Disclosure \\
			\midrule
			MarsDB & Command injection & \textsc{BackREST} & \checkmark & Critical &\ifnoblind\cite{marsdb}\else\cite{marsdb-anon}\fi \\
			Sequelize & Denial-of-Service & \textsc{BackREST} & & Moderate & \ifnoblind\cite{sequelize}\else\cite{sequelize-anon}\fi \\
			Apostrophe & Denial-of-Service & \textsc{BackREST} & & --- &\ifnoblind\cite{apostrophe1}\else\cite{apostrophe1-anon}\fi \\
			Apostrophe & Denial-of-Service & \textsc{BackREST} & & Low & \ifnoblind\cite{apostrophe2}\else\cite{apostrophe2-anon}\fi \\
			Mongo-express$^\dagger$ & Command injection & \textsc{BackREST} & \checkmark & Critical & \cite{mongo-express} \\
			Mongo-express & Denial-of-Service & \textsc{BackREST}, Zap, Arachni, w3af & & Medium & \ifnoblind\cite{mongo-express2}\else\cite{mongo-express2-anon}\fi \\
			Mongo-express & Denial-of-Service & \textsc{BackREST}, Zap, Arachni, w3af & & TBA & TBA \\
			Mongo-express & Denial-of-Service & Zap, Arachni, w3af & & TBA & TBA \\
			Mongodb-query-parser & Command injection & \textsc{BackREST} & \checkmark & Critical & \ifnoblind\cite{mongodb-query-parser} \else\cite{mongodb-query-parser-anon}\fi \\
			MongoDB & Denial-of-service & \textsc{BackREST}, Zap & & High & \ifnoblind\cite{mongodb}\else\cite{mongodb-anon}\fi \\
			\bottomrule
		\end{tabular}
		\label{tab:0days}
		\begin{tablenotes}
			\small
			\item $^\dagger$ \textsc{BackREST} independently and concurrently found the same vulnerability
		\end{tablenotes}
	\end{threeparttable}
\end{table*}

\subsection{Reported 0-days}

\autoref{tab:0days} shows the 0-days that were identified in
benchmark applications and their libraries. The \emph{taint only} column shows
whether a particular 0-day was reported with taint feedback \emph{only}. Out of all the 
vulnerabilities reported in \autoref{tab:fuzzer_bugs}, nine translated into 0-days, out 
of which six were reported by \textsc{BackREST} only. Several reasons explain why not
all vulnerabilities translated to 0-day. First, recall that Nodegoat and 
Juice-Shop are deliberately insecure applications with seeded vulnerabilities. While 
\textsc{BackREST} detected several of them, they are not 0-days. Interestingly, 
however, \textsc{BackREST} did report non-seeded vulnerabilities in MarsDB, and
Sequelize, which happen to be dependencies of Juice-Shop. Through the fuzzing of Juice-Shop, 
\textsc{BackREST} indeed triggered a command injection vulnerability in MarsDB and a 
denial-of-service in Sequelize. Second, the XSS that were reported in 
Apostrophe and Keystone are exploitable only in cases where the JSON response 
containing the XSS payload is processed and rendered in HTML. While we argue that
returning JSON objects containing XSS payloads is a dangerous practice, developers 
decided otherwise and did not accept our reports as vulnerabilities. Third, 
Mongo-express is a database management console; it 
deliberately lets its users inject arbitrary content. Hence, NoSQLi in Mongo-express 
can be considered as a \emph{feature}. Otherwise, the command injection and denial-of-service vulnerabilities
in Mongo-express and its dependencies all translated into 0-days, and so did the 
denials-of-services in Apostrophe.

For readers who might not be familiar with the Node.js ecosystem, it is important to 
underline how the MongoDB and Sequelize libraries are core to \emph{millions} of
Node.js applications. At the time of writing, MongoDB\footnote{\url{https://www.npmjs.com/package/mongodb}} had 1\,671\,653 \emph{weekly} downloads while Sequelize\footnote{\url{https://www.npmjs.com/package/sequelize}} had 648\,745. By any standard, these libraries are extremely heavily used, and well exercised. 

\section{Case studies}
\label{sec:case-studies}

\begin{lstlisting}[style=JavaScript, caption={Command injection vulnerability in \texttt{MarsDB}}, label=lst:marsdb-vuln, float=*],
//Juice-Shop code
//Implements the /rest/track-order/{id} route
db.orders.find({ $where: "this.orderId === '" + (*\textbf{req.params.id}*) + "'" }).then(
  order => { ... }, 
  err => { ... }
);

//MarsDB code
$where: function((*\textbf{selectorValue}*), matcher) {
  matcher._recordPathUsed('');
  matcher._hasWhere = true;
  if (!(selectorValue instanceof Function)) {
    selectorValue = Function('obj', 'return ' + (*\textbf{selectorValue}*));
  }
  return function(doc) {
    return {result: (*\textbf{selectorValue}*).call(doc, doc)};
  }
};
\end{lstlisting}

In this section, we detail some of the 0-days we reported in \autoref{tab:0days}.
We also explain some JavaScript constructs that might be puzzling to readers who
are not familiar with the language. All the information presented in the following
case studies is publicly available and a fix has been released for all but one of
the vulnerabilities we present (MarsDB). In this particular case, the vulnerability
report has been public since Nov 5th, 2019.

\subsection{MarsDB command injection}

MarsDB is an in-memory database that implements the MongoDB API.
\autoref{lst:marsdb-vuln} shows the command injection vulnerability in the 
MarsDB library that \textsc{BackREST} uncovered.  Attacker-controlled 
input is injected in the client application at line 3, through a request parameter (bolded). 
The client application then uses the unsanitised tainted input to build a MarsDB
\lstinline[style=JavaScript]|find| query. In Node.js, long-running operations, such as
querying a database, are executed asynchronously. In this example, calling the \lstinline[style=JavaScript]|find| method returns a JavaScript \emph{promise} that
will be resolved asynchronously. Calling the \lstinline[style=JavaScript]|then| method
of a promise allows to register handler functions for cases where the promise is fulfilled
(line 4) or rejected (line 6). The query eventually reaches the 
\lstinline[style=JavaScript]{where} function of the MarsDB library at line 9 as the \lstinline[style=JavaScript]{selectorValue} argument. That argument is then used 
at line 13 to dynamically create a new function from a string. From a security perspective,
calling the \lstinline[style=JavaScript]|Function| constructor in JavaScript is 
roughly equivalent to calling the infamous \lstinline[style=JavaScript]|eval|; it 
dynamically creates a function from code supplied as a string. The newly created function 
is then called at line 16, which triggers the command injection vulnerability \ifnoblind\cite{marsdb} \else\cite{marsdb-anon} \fi. 
In this particular case,
unless the payload is specifically crafted to: 1. generate a string that is valid JavaScript
code, and 2. induce a side-effect that is observable from the client, it can be very difficult to detect
this vulnerability in a purely blackbox manner. Thanks to taint feedback, \textsc{BackREST}
can detect the command injection as soon as \emph{any} unsanitised input reaches the \lstinline[style=JavaScript]|Function| constructor at line 16.

\subsection{Apostrophe DoS}

\begin{lstlisting}[style=JavaScript, caption={Denial-of-Service (DoS) vulnerability in \texttt{Apostrophe}}, label=lst:apos-vuln, float=*],
self.routes.list = function(req, res) {
  ...
  if (req.body.format === 'managePage') {
    ...
  } else if (req.body.format === 'allIds') {
    ...
  }
  return self.listResponse(req, res, err, results);
};
\end{lstlisting}

Apostrophe is an enterprise content management system (CMS). \autoref{lst:apos-vuln}
shows a snippet of Apostrophe code that is vulnerable to a DoS attack. This code reads
the \lstinline[style=JavaScript]|format| parameter of the request body, and checks if it
is equal to ``\texttt{managePage}'' or ``\texttt{allIds}'', but misses a fallback
option for cases where it is equal to neither. If this situation occurs, an uncaught
exception is thrown, crashing the server \ifnoblind\cite{apostrophe1} \else\cite{apostrophe1-anon} \fi. 

\subsection{Mongo-express command injections}

Mongo-express is a MongoDB database management console. In versions prior to 0.54.0, it
was calling an eval-like method with attacker-controllable input, leading to a command injection
vulnerability. While \textsc{BackREST} independently detected this vulnerability, it was
concurrently reported days before our own disclosure \cite{mongo-express}. Interestingly,
\textsc{BackREST} then revealed how the fix still enabled command injection. Indeed, 
the fix was to use the \texttt{mongo-db-query-parser} library to parse attacker-controlled input. 
The issue is that the library is using \lstinline[style=JavaScript]|eval| itself. Thanks to taint 
feedback, \textsc{BackREST} detected that tainted input was \emph{still} flowing to an 
\lstinline[style=JavaScript]|eval| call, which we disclosed \ifnoblind\cite{mongodb-query-parser} \else\cite{mongodb-query-parser-anon} \fi.

\subsection{Sequelize DoS}

\begin{lstlisting}[style=JavaScript, caption={Denial-of-Service (DoS) vulnerability in \texttt{Sequelize}}, label=lst:sequelize-vuln, float=*],
run((*\textbf{sql}*), parameters) {
  (*\textbf{this.sql}*) = sql;
  ...
  const (*\textbf{query}*) = this;
  function afterExecute(err, results) {
  ...
    if ((*\textbf{query.sql}*).indexOf('sqlite_master') !== -1) {
      if ((*\textbf{query.sql}*).indexOf('SELECT sql FROM sqlite_master WHERE tbl_name') !== -1) {
        result = results;
        if (result && result[0] && result[0].sql.indexOf('CONSTRAINT') !== -1) {
          result = query.parseConstraintsFromSql(results[0].sql);
        }
      } 
      else if (results !== undefined) {
        // Throws a TypeError if results is not an array.
        result = results.map(resultSet => resultSet.name);
      } 
      else { 
        result = {}
      }
    }
  ...
  }
}
\end{lstlisting}

Sequelize is a Node.js Object-Relational Mapper (ORM) for Postgres, MySQL, MariaDB, 
SQLite and Microsoft SQL Server. The code in \autoref{lst:sequelize-vuln} is vulnerable
to a DoS attack that crashes the Node.js server with an uncaught exception. First, an
attacker-controllable SQL query is passed as the \lstinline[style=JavaScript]|sql| argument
to the \lstinline[style=JavaScript]|run| function that executes SQL queries at line 1.
The tainted query is assigned to various variables, the query is executed, and after its
execution is eventually searched for the string ``\texttt{sqlite\_master}'' at line 7, 
which is a special table in SQLite. If the search is successful, the query is then searched 
for the string ``\texttt{SELECT sql FROM sqlite\_master WHERE tbl\_name}''. If this search is unsuccessful and the query returned a \lstinline[style=JavaScript]|results| object that is
not \lstinline[style=JavaScript]|undefined| (line 14), the \lstinline[style=JavaScript]|map|
\footnote{\url{https://developer.mozilla.org/en-US/docs/Web/JavaScript/Reference/Global_Objects/Array/map}} method is called on the \lstinline[style=JavaScript]|results| object at line 16.
This is where the DoS vulnerability lies. If the \lstinline[style=JavaScript]|results|
object is not an array, it likely won't have a \lstinline[style=JavaScript]|map| method in
its prototype chain, which will throw an uncaught \texttt{TypeError} and crash the Node.js process 
\ifnoblind\cite{sequelize} \else\cite{sequelize-anon} \fi. 
In summary, any request that includes the string ``\texttt{sqlite\_master}'', 
but not ``\texttt{SELECT sql FROM sqlite\_master WHERE tbl\_name}'', and that returns a 
single value (i.e. not an array), will crash the underlying Node.js process.

\subsection{MongoDB DoS}

\begin{lstlisting}[style=JavaScript, caption={Denial-of-Service (DoS) vulnerability in \texttt{MongoDB}}, label=lst:mongodb-vuln, float=*htb],
function createCollection(db, (*\textbf{name}*), options, callback) {
  ...
  executeCommand(db, cmd, finalOptions, 
    err => {
      if (err) return handleCallback(callback, err);
      handleCallback(
        callback,
        null,
        // Throws an uncaught MongoError if the name argument is invalid
        new Collection(db, db.s.topology, db.s.databaseName, (*\textbf{name}*), db.s.pkFactory, options)
      );
    }
  );
  ...
}
\end{lstlisting}

MongoDB is a document-based NoSQL database with drivers in several languages.
\autoref{lst:mongodb-vuln} shows a snippet from the MongoDB driver that has
a DoS vulnerability. This code gets executed when new collections are created in a
MongoDB database. If the name of the collection to be created is attacker-controllable,
and the attacker supplies an invalid collection name, the
call to the \lstinline[style=JavaScript]|Collection| constructor at line 10 fails
and throws an uncaught \lstinline[style=JavaScript]|MongoError| that crashes the 
Node.js process \ifnoblind\cite{mongodb} \else\cite{mongodb-anon} \fi. The taint feedback loop quickly reports
that a tainted collection name flows to the \lstinline[style=JavaScript]|Collection|
constructor, enabling \textsc{BackREST} to trigger the vulnerability faster.

\section{Related Work}
\label{sec:relwork}

\paragraph{Web application modelling}

Modelling for web applications has a long and rich history in
the software engineering and testing communities. Modelling methods broadly fall 
into three main categories: graph-based, UML-based, and FSM-based. Graph-based 
approaches focus on extracting navigational graphs from web applications and
applying graph-based algorithms (e.g. strongly connected components, dominators)
to gain a better understanding of the application \cite{tonella2002dynamic,di2002anweb}. 
UML-based approaches further capture the interactions and flows of data between 
the different components of a web application (e.g. web pages, databases, and 
server) as a UML model \cite{ricca2001analysis, tonella2002dynamic, antoniol2004understanding}.
To facilitate automated test case generation, FSM-based approaches instead cluster 
an application into sub-systems, model each with a finite-state machine and unify 
them into a hierarchy of FSMs \cite{andrews2005testing}. All these approaches were
designed to model \emph{stateful} web applications, which were the norm back in the
early 2000s. Since then, web development practices evolved, developers realised
that building server-side applications that are as \emph{stateless} as possible
improves maintainability, and the REST protocol, which encourages statelessness, gained 
significant popularity.

\paragraph{Grammar inference} 
Automated learning of grammars from inputs is a 
complementary and very promising research area \cite{godefroid2017,hoschele2017,bastani2017synthesizing}. 
To efficiently learn a grammar from inputs, however, current approaches
either require: 1. very large datasets of input to learn from (\cite{godefroid2017}); 
2. highly-structured parser code that reflects the structure of the underlying 
grammar (\cite{hoschele2017}); or 3. a reliable oracle to determine whether a given
input is well-formed (\cite{bastani2017synthesizing}). Unfortunately, very few web 
applications meet any of these criteria, making model inference the only
viable alternative. Compared to synthesised grammars, REST models are also 
easier to interpret.

\paragraph{Model-based fuzzing}

Model-based fuzzing derives input from a user-supplied \cite{pham2016model,atlidakis2019restler}, 
or inferred
 \cite{schneider2013online, schneider2012behavioral, han2017imf, duchene2014kameleonfuzz, doupe2012enemy} model. 
Contrary to grammar-based fuzzing \cite{holler2012fuzzing, hoschele2016mining,	godefroid2008grammar}, input generation in model-based fuzzing is not constrained by a context-free grammar. While grammar-based fuzzers generally excel at fuzzing language 
parsers, model-based fuzzers are often better suited for higher-level programs
with more weakly-structured input.

\paragraph{REST-based fuzzing}

Most closely related to our work are REST-based fuzzers. In recent years, many HTTP fuzzers 
have been extended to support REST specifications \cite{apifuzzer,appspider,tntfuzzer,qualys,swurg},
but received comparatively little attention from the academic community. RESTler \cite{atlidakis2019restler} is the exception. It uses a user-supplied REST 
specification as a model for fuzzing REST-based services in a blackbox manner. To better
handle stateful services, RESTler enriches its model with inferred dependencies 
between endpoints. As we highlighted in \autoref{sec:stateful}, we haven't found endpoint 
dependency inference to have a significant impact on the coverage of all but one of our 
benchmark applications. For Mongo-express, we found that the inferred dependencies were 
trivial, easily configurable and did not justify the added complexity.

\paragraph{Taint-based fuzzing}

Taint-based fuzzing uses dynamic taint analysis to monitor the flow of attacker-controlled
values in the program under test and to guide the fuzzer
 \cite{ganesh2009taint, bekrar2012taint, liang2013effective}. Wang et al. \cite{wang2010taintscope,wang2011checksum} use taint analysis to identify the parts of 
an input that are compared against checksums. Similarly, Haller et al. use taint analysis 
to identify inputs that can trigger buffer overflow vulnerabilities \cite{haller2013dowsing},
while Vuzzer \cite{rawat2017vuzzer} uses it to identify parts of an input that are 
compared against magic bytes or that trigger error code. Taint analysis has also been
used to map input to the parser code that processes it and to infer an input grammar \cite{hoschele2016mining}. More closely related to our work, the KameleonFuzz tool \cite{duchene2014kameleonfuzz} uses taint inference on the client-side of web 
applications to detect reflected inputs and exploit cross-site scripting vulnerabilities.
Black Widow \cite{eriksson2021black} implements several stateful crawling strategies in 
combination with taint inference to detect stored and reflected cross-site scripting 
vulnerabilities in PHP applications.

\paragraph{Web application security scanning}

The process of exercising an application with automatically generated malformed, or
malicious input, which is nowadays known as fuzzing, is not new, and is also known 
as security scanning, attack generation, or vulnerability testing in the web 
community. Whitebox security scanning tools include the QED tool \cite{martin2008automatic} 
that uses goal-directed model checking to generate SQLi and XSS attacks for Java web 
applications. The seminal Ardilla paper \cite{kieyzun2009automatic} presented a whitebox
technique to generate SQLi and XSS attacks using taint analysis, symbolic databases,
and an attack pattern library. Ardilla was implemented using a modified PHP interpreter, 
tying it to a specific version of the PHP runtime. Similarly, Wassermann et al. also 
modified a PHP interpreter to implement a concolic security testing approach 
\cite{wassermann2008dynamic}. \textsc{BackREST} instrumentation-based 
analyses decouples it from the runtime, making it easier to maintain over time.
More recent work on PHP application security scanning includes Chainsaw 
\cite{alhuzali2016chainsaw} and NAVEX \cite{alhuzali2018navex} that use static 
analysis to identify vulnerable paths to sinks and concolic testing to generate concrete
exploits. Unfortunately, the highly dynamic nature of the JavaScript language makes
any kind of static or symbolic analysis extremely difficult. State-of-the-art static 
analysis approaches can now handle some libraries and small applications \cite{stein2019static, 
nielsen2019nodest, ko2019weakly} but concolic testing engines still struggle to handle 
more than a thousand lines of code \cite{dhok2016type, loring2017expose, amadini2019constraint}.
For this reason, blackbox scanners like OWASP Zap \cite{zap}, Arachni \cite{arachni} or
w3af \cite{w3af}, which consist of a crawler coupled with a fuzzing component, were the 
only viable option for security scanning of Node.js web applications. With \textsc{BackREST},
we showed that lightweight coverage and taint inference analyses are well-suited to dynamic
languages for which static analysis is still extremely challenging.

\paragraph{Web vulnerability detection and prevention}

In the past two decades, a very large body of work has focused on detecting and
preventing vulnerabilities in web applications. The seminal paper by Huang et al.
introduced the WebSSARI tool \cite{huang2004securing} that used static analysis to 
detect vulnerabilities and runtime protection to secure potentially vulnerable
areas of a PHP application. In their 2005 paper, Livshits and Lam showed how static
taint analysis could be used to detect injection vulnerabilities, such as SQLi 
and XSS, in Java web applications \cite{livshits2005finding}.
The Pixy tool \cite{jovanovic2006pixy} then showed how static taint analysis could
be ported, to the PHP language to detect web vulnerabilities in
PHP web applications. The AMNESIA tool \cite{halfond2005amnesia} introduced the idea
of modelling SQL queries with static analysis and checking them against the model at
runtime. This idea was further formalised by Su et al. \cite{su2006essence}, applied
to XSS detection \cite{ter2009blueprint} and is still used nowadays to counter 
injection attacks in Node.js applications \cite{staicu2018synode}. 

\paragraph{JavaScript vulnerability detection}

As Web 2.0 technologies gained in
popularity, the client-side of web applications became richer, and researchers
started investigating the JavaScript code that runs in our browsers. It became
quickly obvious, however, that existing static analysis techniques could not be
easily ported to JavaScript, and that dynamic techniques were better suited for
highly dynamic JavaScript code \cite{saxena2010flax}. Dynamic taint analysis
thus started to gain popularity, and was particularly successful at detecting
client-side DOM-based XSS vulnerabilities \cite{lekies201325,parameshwaran2015dexterjs}.
In the meantime, in 2009, the first release of Node.js, which brings JavaScript to
the server-side, came out, and is now powering millions of web applications 
worldwide. Despite its popularity, however, the Node.js platform comparatively 
received little attention from the security community \cite{ojamaa2012assessing, pfretzschner2017identification, davis2017node}, with only two studies addressing
injection vulnerabilities \cite{staicu2018synode,nielsen2019nodest}.

%
%

\section{Conclusion}
\label{sec:conclusion}

We presented \textsc{BackREST}, the first fully automated model-based,
coverage- and taint-driven greybox fuzzer for web applications. \textsc{BackREST}
guides a state-aware crawler to automatically infer REST APIs, and uses coverage feedback 
to avoid fuzzing thoroughly covered code. \textsc{BackREST} makes a novel use of taint 
feedback to focus the fuzzing session on likely vulnerable areas, 
guide payload generation, and detect more vulnerabilities. Compared to a baseline
version without taint and coverage feedback, \textsc{BackREST} achieved speedups 
ranging from 7.4$\times$ to 25.9$\times$. \textsc{BackREST} also consistently 
detected more (No)SQLi, command injection, and XSS vulnerabilities than three state-of-the-art 
web fuzzers and detected six 0-days that were missed by all other fuzzers.

By extending a blackbox web application fuzzer with lightweight whitebox analysis, 
we showed how greybox fuzzing can detect \emph{more} vulnerabilities \emph{faster}.
We hope that our study will trigger further research in this area and encourage
researchers to develop lightweight greybox fuzzers for web applications written in 
languages that have received little attention so far, like Python and Ruby.

\bibliographystyle{IEEEtran}
\bibliography{refs}

\end{document}